\documentclass{article}
\usepackage{graphicx}
\usepackage{subfigure}
\usepackage{amssymb}
\usepackage{float}
\usepackage{wrapfig}
\usepackage{listings}
\usepackage{adjustbox,lipsum}
\makeatletter
\renewcommand\part{%
  \if@openright
    \cleardoublepage
  \else
    \clearpage
  \fi
  \thispagestyle{empty}%
  \if@twocolumn
    \onecolumn
    \@tempswatrue
  \else
    \@tempswafalse
  \fi
  \null\vfil
  \secdef\@part\@spart}
\makeatother
\usepackage[numbers,sort]{natbib}
\usepackage{authblk}
\usepackage[pdfborder={0 0 0}]{hyperref}

\usepackage{mwe,tikz}
\usepackage[percent]{overpic}
\usepackage{tikz}
\usepackage{tkz-tab}
\usepackage{array}
\newcolumntype{C}[1]{>{\centering\arraybackslash}p{#1}}

\begin{document}
%
\title{Impact of the spatial context on human communication activity}

\author[1]{Zolzaya Dashdorj}
\author[2]{Stanislav Sobolevsky}

\affil[1]{University of Trento and SKIL LAB - Telecom Italia and DKM - Fondazione Bruno Kessler, Italy Via Sommarive, 9 Povo, TN, Italy}
\affil[ ]{\url{dashdorj@disi.unitn.it}}
\affil[2]{Massachusetts Institute of Technology, MIT 77 Massachusetts Avenue Cambridge, MA, USA}
\affil[ ]{\url{stanly@mit.edu}}


%


\date{}
\maketitle

\begin{abstract}
Technology development produces terabytes of data generated by human activity in space and time. This enormous amount of data often called big data becomes crucial for delivering new insights to decision makers. It contains behavioral information on different types of human activity influenced by many external factors such as geographic information and weather forecast. Early recognition and prediction of those human behaviors are of great importance in many societal applications like health-care, risk management and urban planning, etc. In this paper, we investigate relevant geographical areas based on their categories of human activities (i.e., working and shopping) which identified from geographic information (i.e., Openstreetmap). We use spectral clustering followed by k-means clustering algorithm based on TF/IDF cosine similarity metric. We evaluate the quality of those observed clusters with the use of silhouette coefficients which are estimated based on the similarities of the mobile communication activity temporal patterns. The area clusters are further used to explain typical or exceptional communication activities. We demonstrate the study using a real dataset containing 1 million Call Detailed Records. This type of analysis and its application are important for analyzing the dependency of human behaviors from the external factors and hidden relationships and unknown correlations and other useful information that can support decision-making. 
\end{abstract}

\begin{keywords}
telecommunication dataset, human behavior, cell phone data records, activity recognition, knowledge management, clustering and classification
\end{keywords}


\section{Introduction}\label{introduction}

Nowadays extensive penetration of digital
technologies into everyday life creates vast amount of data
related to different types of human activity. When
available for the research purposes this creates an
unprecedented opportunity for understanding human
society directly from it’s digital traces. There is an impressive amount of papers leveraging such data for studying human behavior, including mobile phone records \cite{ratti2006mlu, calabrese2006real, girardin2008digital, quercia2010rse, ratti2010redrawing,amini2013differing}, vehicle GPS traces \cite{santi2014quantifying, kang2013exploring}, social media posts \cite{java2007we, hawelka2014, paldino2015urban} and bank card transactions \cite{sobolevsky2014mining, sobolevsky2014money,sobolevsky2015scaling}. And mobile phone data is among the most commonly used data sources from above. Such data allows us to understand human behaviors and social relationships by investigating the influence of context factors of social dynamics. Potential applications of this research are a context aware application systems, and  ``smart cities'' applications that provide decision support for stakeholders in areas such as urban, transport planning, tourism and event analysis, emergency response, health improvement, community understanding, economic indicators and others. Current research \cite{DBLP:journals/corr/abs-1106-0560, Phithakkitnukoon:2010:AMI:1881331.1881336, Calabrese:2010:GTA:2166616.2166619,Furletti:2012:IUP:2346496.2346500,DBLP:conf/mum/DashdorjSAL13} notes that with a pure CDR, it is possible to identify human behaviors, but results suffer from the heterogeneity, uncertainty and complexity of raw datasets and that the lack of qualitative content is included in the data itself that may be used to help to infer human behaviors. \cite{Zolzaya2015} identifies that Points of Interest (POIs) provide a good proxy for predicting the content of human activities in each area and thus for identifying the activities people are more likely to perform. This is much more effective if we combine mobile phone data records that  can be more likely associated to human activities, for example, a person looking for some food if phone call is located in or close to a restaurant. In this paper, we concentrated on characterization and clustering of geographical areas based on the categories of human activity in order to identify relevant areas. The model proposed in \cite{DBLP:conf/mum/DashdorjSAL13}, is used to extract top level human activities from geographical open data source. We use spectral clustering with eigengap heuristic followed by k-means clustering and intrinsic method to evaluate the quality of clusters. In each area cluster, we contextually enrich the mobile phone data records with the categories of human activities and then analyze and identify the standard or exceptional (divergent) type of the communication activity temporal patterns. This further opens a discussion to understand various type of relationships between environment and human behavior. The paper is structured as follows Section \ref{related works} illustrates the related works and methodology is described in Section \ref{methodology}. We present and discuss the results in Section \ref{results and discussion}. Finally, we summarize the discussions in Section \ref{conclusion}.

\section{Related works}\label{related works}

The clustering approaches ( Han \& Kamber \cite{citeulike:2855241}) such as k-means, k-medoids, and self organizing map group similar spatial objects into classes, and several other methods are also used to perform effective and efficient clustering, for instance, Calabrese et al \cite{RePEc:pio:envirb:v:36:y:2009:i:5:p:824-836} and also \cite{pei2014new,DBLP:journals/corr/GrauwinSMGR14} used eigengap heuristic for clustering. Phithakkitnukoon et al \cite{Phithakkitnukoon:2010:AMI:1881331.1881336} identified area profile from POIs. Each area is connected to main activity considering the category of POIs, and activity patterns for each mobile user are studied to determine groups which have similar activity patterns. Noulas et al \cite{DBLP:conf/icwsm/NoulasSMP11a} proposed an approach for modeling and characterization of geographic areas based on a number of user check-ins and a set of 8 general (human) activity categories in Foursquare. Cosine similarity metric is used to measure a similarity of geographical areas, and spectral clustering algorithm that followed by k-means clustering is applied to identify a relevant area profile. The area profiles enables to understand  groups of individuals who have similar activity patterns. Similar to this research idea, social networks\cite{Wakamiya:2011:UAC:2008664.2008674} have been taken into account to discover activity patterns of individuals. Frias-Martinez et al \cite{DBLP:conf/socialcom/Frias-MartinezSHF12} studied geolocated tweets to characterize urban landscapes using a complimentary source of land-use and landmark information. The author focused on determining the land-uses in a specific urban area based on tweeting patterns, and identification of POIs in high activity tweeted areas. Differently, Yuang et al \cite{conf/giscience/YuanR12} proposed to classify urban areas based on their mobility patterns by measuring the similarity between the time-series using Dynamic Time Warping (DTW) algorithm. Some of areas focus on understanding urban dynamics including dense area detection and their evolution over time \cite{Vieira:2010:CDU:1906497.1907403,conf/icde/NiR07}.  


\section{Data-source collection}
We model contextual information of 4.6 million POIs in Trento, Italy and behavioral dataset of 1 million mobile phone data records (CDR). 

\subsection{Openstreetmap}

For modeling the contextual description of geographical regions in Trento, we use a High level Representation of Behavior Model (HRBModel) \cite{Zolzaya2015,DBLP:conf/mum/DashdorjSAL13} which exploits a spatial grid (i.e.,  cell size is 50m x 50m) which populated the locations (i.e., cell) with POIs from open geographic information, Openstreetmap (OSM). The model generates human activity distribution map which enriched by the categories of human activity associated with a likelihood measure. For example, watching a football match on a stadium, eating in restaurant or hiking in forest. We collected in total 135,918 relevant POIs extracted from OSM. After cleaning and discarding irrelevant POIs (i.e., those that do not reflect relevant human activity), the number of POIs is reduced to 31,514 POIs. The total number of human activities are extracted up to 78,068 belongs to the POIs. The top-level classes of activity categories belong to the POIs are explained in Table \ref{table:group action}. 
 
\begin{table}
\centering
    \caption{The categories of human activity belong to the POIs}
	\label{table:group action}
  \resizebox{10cm}{!}{
    \begin{tabular}{|l|l|}
        \hline
        \textbf{Top level classes of activities} & \textbf{Type of POIs} \\ \hline
        eating & fast food, food court, restaurant, cafe\\ \hline
        shopping & grocery, general stores\\ \hline
        health medicine activity & hospital, pharmacy\\ \hline
        entertainment activity & bar, casino, movie, theater\\ \hline
        education activity & library, university school\\ \hline
        transportation traveling & airplane, bus, car, train\\  \hline
        outdoor activity & sightseeing, personal care, religious places\\ \hline
        sporting activity & car racing, summer, winter sports\\ \hline
        working activity & professional work place, industrial place\\ \hline
        residential activity & guest house, hotel, hostel, residential building\\ \hline
    \end{tabular}
   }
\end{table}

\subsection{Mobile phone data records}
We collected CDR about outgoing logged calls for 2 months. The CDR is completely anonymized containing cell ID, time of day and duration in which a phone call is issued. Cell-ID is used for identification of some portion of a physical geographic area featured with a set of devices (antennas) that support the communication. Sometimes, cell coverage area is not precisely defined, and can be temporarily modified depending on the estimation of call traffic from/to this area. Usually the size of the coverage area is inversely proportional to the density of the population inhabiting the area. It is observed that, in presence of regular territory (i.e., flat with no mountains or other natural irregularities), the shape of cells can be approximated with convex polygons, otherwise the cell can be very irregular and possibly disconnected.

\section{Methodology}\label{methodology}

We present our approach for clustering algorithm to identify relevant areas in terms of geographical area profiles containing a set of the categories of human activity and use the observed clusters for analyzing mobile phone communication activities. Our approach is aimed at answering the following questions: What are the geographical area profiles defined by human activities in a city? How is the communication pattern affected by the profile of area activity? To do that, we first identify relevant area clusters based in terms of the category of activities and then analyze the communication activity patterns in those observed area clusters.

\subsection{Geographical area clusters}

We define a vector space model that contains a set of activity categories corresponding to geographical areas. The representation of geographic areas $l_i$ within the territory of a city or large square area $L$. Each area $l_i$ contains a vector of different top level activities derived from the POIs in such area that would be an input data-point for identifying the relevant areas by cluster algorithms. The area features are represented by a matrix $l_{i,j}$ containing the weight of the activity categories $j$ in each area $L_i$. The relevance between the areas is identified by the cosine similarity metric by estimating the deviation of angles among area vectors. For example, the similarity between area $l_1$ and $l_2$ is as $\cos{\theta}_{1,2} = \frac{\mathbf{l_1} \cdot \mathbf{l_2}}{\left\| \mathbf{l_2} \right\| \left \| \mathbf{l_1} \right\|}$. Having the estimation of similarity between the areas, we can now create a similarity graph described as the weight matrix $W$ and the degree matrix $D$ is utilized by the spectral clustering algorithm which is the one of the most popular modern clustering methods and performs better than traditional clustering algorithms. The K-Nearest Neighbors of each data point are identified using cosine similarity metric, we create the adjacency matrix of the similarity graph and graph Laplacian $L=D-A$ (given by normalized graph Laplacian $L_n=D^{-1/2}LD^{-1/2}$). Based on eigengap heuristic, we identify the number of clusters to observe in our dataset as $k = argmax_{i}(\lambda_{i+1}-\lambda_{i})$ where $\lambda_i \in \{l_1, l_2, l_3,.., l_n\}$ denotes the eigenvalues of $L_n$ in the ascending order. Finally, we easily detect the effective clusters (area profiles) $C_1, C_2, C_3, ..., C_k$ from the first $k$ eigenvectors identified by the k-means algorithms.

%
%
%

\subsection{Behavioral pattern extraction in area clusters}
\begin{figure}[h!]
  \centering
  \subfigure[The density of communication activity distribution over typical day]
  {\includegraphics[width=0.3\linewidth]{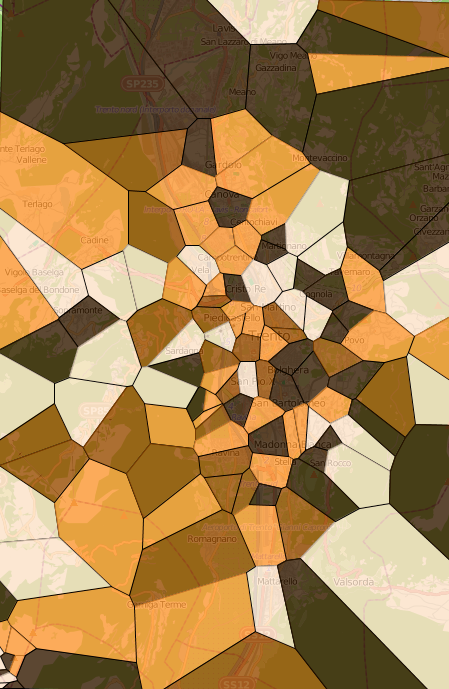}
   \label{gra:distribution calling activity in Normal}}
  \quad
  \subfigure[The density of communication activity distribution over Easter Sunday]
  {\includegraphics[width=0.3\linewidth]{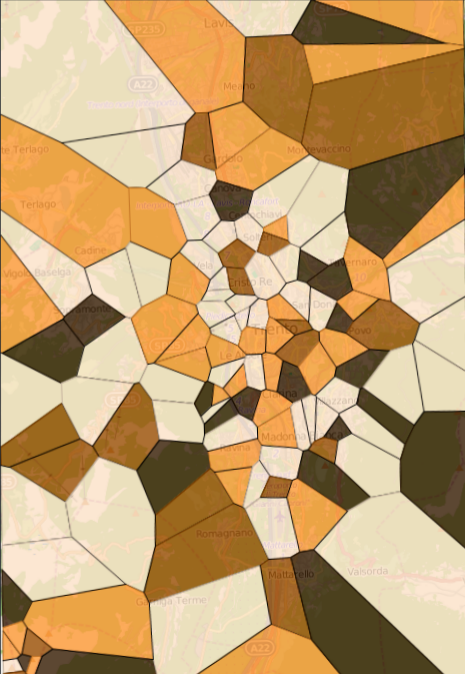}
  \label{gra:distribution calling activity in Easter}}
  \caption{The density of communication activity distribution is represented in the colors black belongs to high volume activity and white belongs low volume activity}
\end{figure}

We are interested in the contextualization of communication activity temporal patterns in relevant area clusters in order to determine standard or exceptional type of communication activities based on the communication activity variations in different time context. This could be done by overlapping between human activity distribution map and cell coverage map. However in this analysis, the cell coverage map is unavailable and the location of cell towers are given approximately. We decided to divide the study area into Voronoi polygons based on the spatial distribution of cell phone towers. The Figure \ref{gra:distribution calling activity in Normal} shows the Voronoi polygons for visualizing cell coverage map. We are then  able to extract mobile communication activities in each polygon as a coverage area. For extracting mobile communication activities in observed area clusters, we need to associate each Voronoi polygon to the human activity distribution map. A given cell area $p$ might be intersected with multiple areas $l_i$ that represented as a list of areas with an intersection weight [0,1]. The intersection weight is estimated by the division of the areas size $l_i$ and $p$, as $W(p,l_i) = \frac{S(l_i)}{S(p)}$. The number of calls per area is the number of calls in a given cell $p$ at a certain time $t$ divided by the number of intersecting areas $N$, taking the intersection weights into account, as $X(p,t,l_i) = \frac{X(p,t)}{N} \cdot w_i$.

We are now able to extract mobile communication activity patterns in different area clusters. In order to identify a normal (typical) type of communication activity temporal pattern, we exclude the communication activities over specific days when the public holidays or festivals occur, such as Liberation day, Palm Sunday and Easter Holiday (see the example during Easter holiday, Figure \ref{gra:distribution calling activity in Easter}) because the communication activities could have significant changes. To estimate the variation boundaries of the typical communication activities over different time context in a given area cluster we use a sigma approach in order to determine exceptional (divergent) communication activity temporal patterns in each area cluster, as $X'(c,t)= \mu_{X(c,t)} \pm \alpha \cdot \sigma_{X(c,t)}$.

\section{Experimental Results and Discussion}\label{results and discussion}

\begin{figure}
  \centering
  \includegraphics[width=0.6\linewidth]{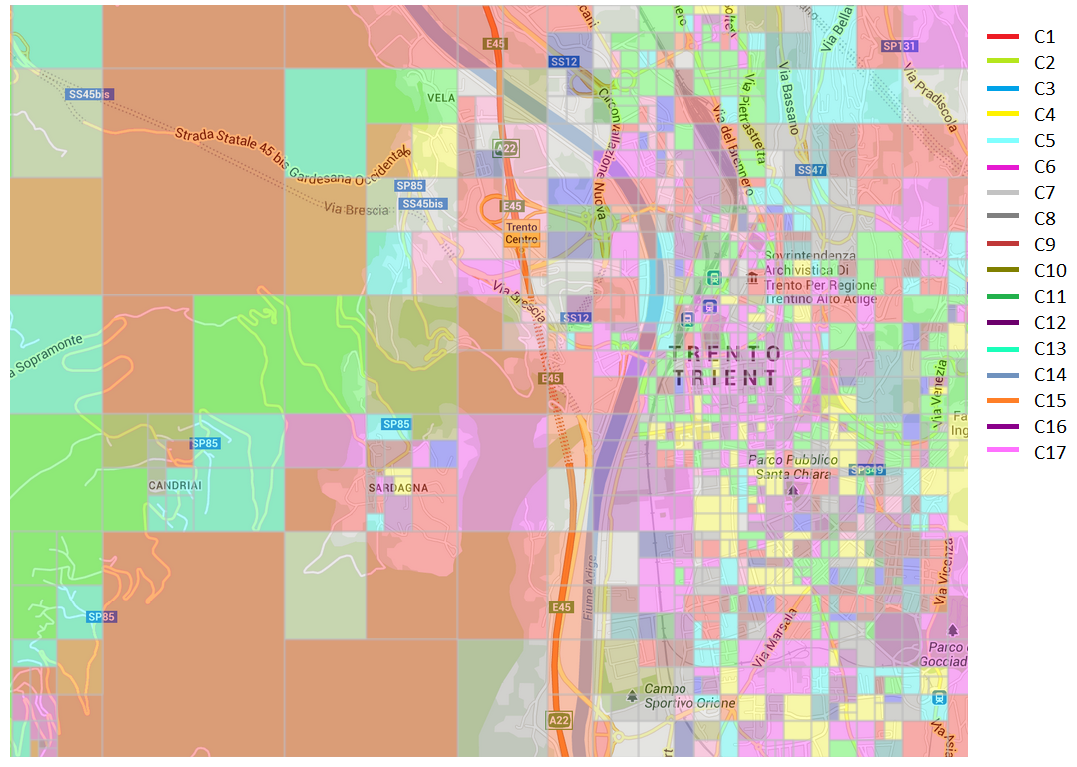}
  \caption{Geo-visualization of area classifications ($k$=17), Trento, Italy}
  \label{gra:visualization cluster}
\end{figure}

\begin{figure}
  \centering
  \includegraphics[width=0.7\linewidth]{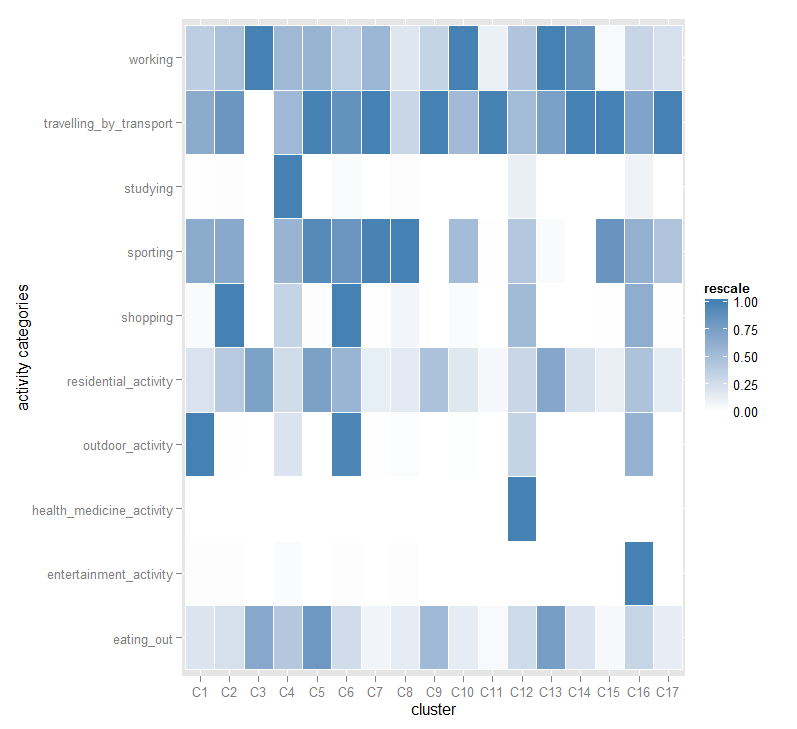}
  \caption{The scaled weight of activity categories of each cluster}
  \label{gra:heatmap clusters}
\end{figure}

\begin{figure}[htb!] \centering
  \includegraphics[width=0.73\linewidth]{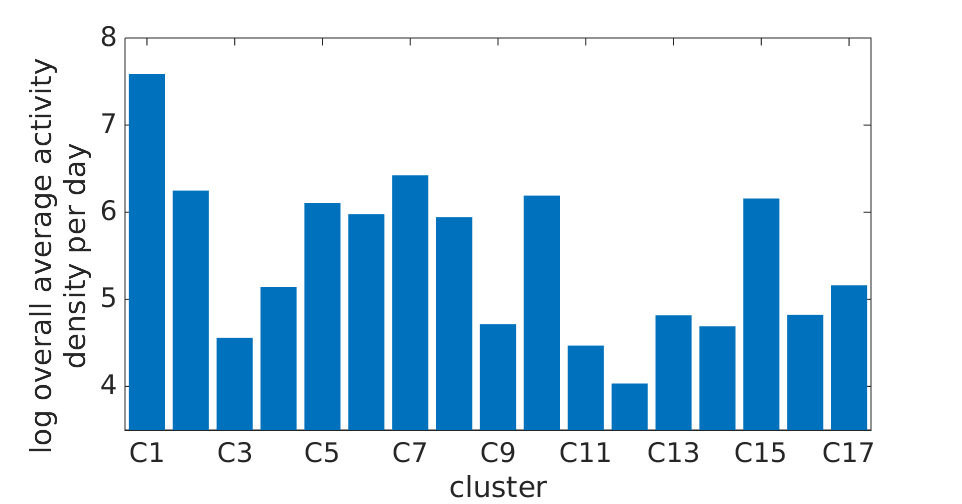}
  \caption{Overall average communication activity density (log scaled) per day with the actual values}
  \label{gra:overall activity density per cluster}
\end{figure}

\begin{figure*}[htb!]
\centering
  \resizebox{10cm}{!}{
	\begin{tikzpicture}[      
        every node/.style={anchor=south west,inner sep=0pt},
        x=3mm, y=4mm,
      ]   
     \node (fig1) at (0,0)
       {\includegraphics[scale=0.25]{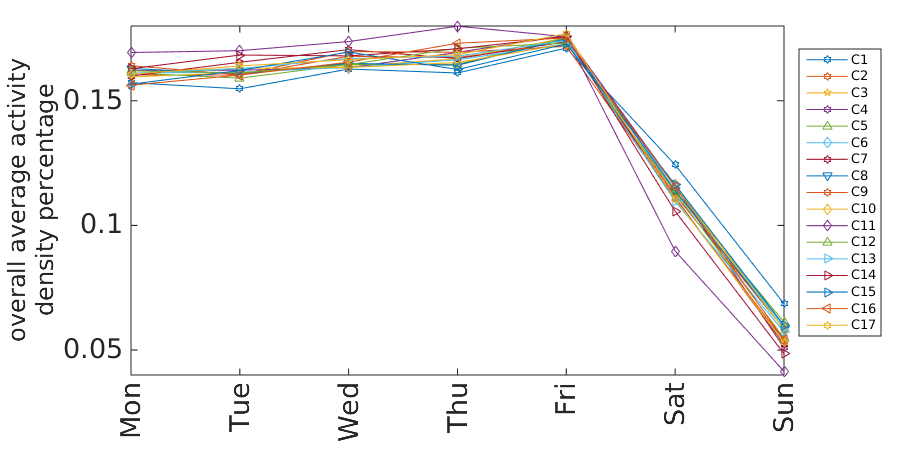}};
     \node (fig2) at (3.5,1.5)
       {\includegraphics[scale=0.12]{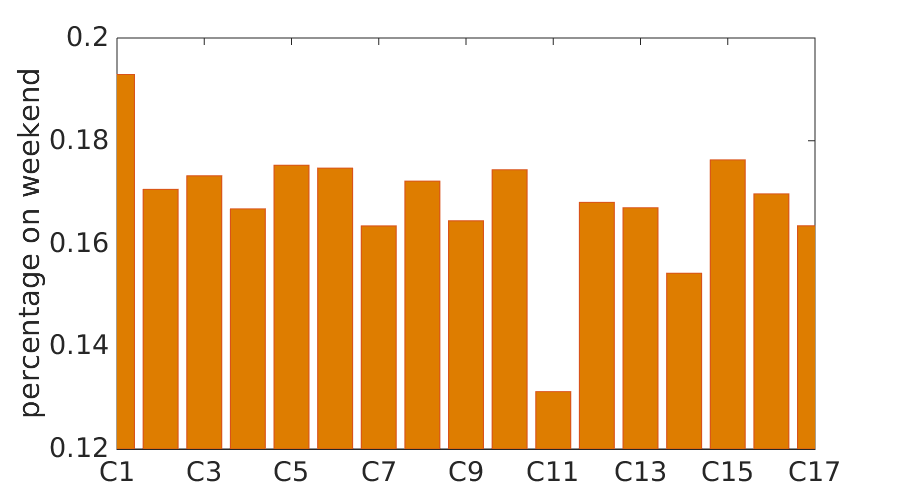}};  
	\end{tikzpicture}}
 \caption{Communication activity temporal variations on the day of week and the subplot is about overall average communication activity density over weekend}
 \label{gra:temporal variations weekly}
\end{figure*}

We are concentrated in identifying relevant areas in which we show how semantics of human activities could be observed to interpret standard or exceptional communication activity temporal patterns. We observed 17 clusters (area profiles) according to activity vector of each area as shown in Figure \ref{gra:visualization cluster}. For each cluster, we show the weight vector of customer activities for each cluster as described in Figure \ref{gra:heatmap clusters}. The central part of the city is clustered into $C_6$, $C_{2}$ followed by $C_{12}$, $C_{16}$ where entertainment, residential, shopping, sporting and traveling by transport activities are highly distributed. We then extract the communication activity temporal patterns in order to see how communication behavior is affected by types of customer activity. The overall average communication activity density per day for each cluster varies depending on the category of the customer activities as shown in Figure \ref{gra:overall activity density per cluster}. $C_{1}$ is the most active cluster in terms of mobile phone activity communication. The weekly temporal communication activity variations per cluster are showing behaviors, similar to each other, see Figure \ref{gra:temporal variations weekly}.

Also there is clearly one purple cluster $C_{11}$ (traveling by transport) which is more active over the weekdays compared to other clusters and less over the weekends and another one light-blue $C_{1}$ (outdoor activity), which shows the opposite pattern to $C_{11}$. This patterns are highlighted in the subplot of Figure \ref{gra:temporal variations weekly} where the total percentage of weekend activity is reported. The different clusters patterns are different in terms of time context (e.g., hour of a day and day of a week). 

\begin{figure}[p]
  \centering
  \subfigure[Daily communication activity pattern per cluster over weekday. The cluster patterns are more diverse over weekend while the patterns are almost similar over weekdays]
  {\includegraphics[width=0.7\linewidth]{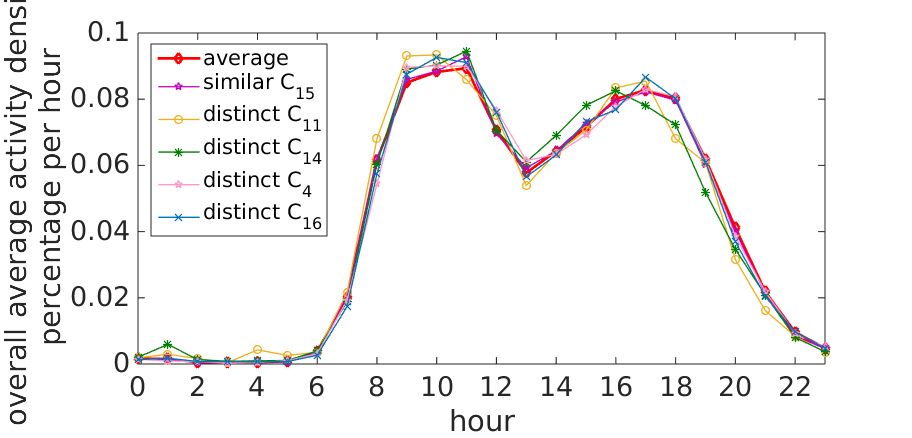}
  \label{gra:daily calling pattern weekday}}
  \quad
  \subfigure[Daily communication activity pattern per cluster over Saturday. In the pattern $C_{11}$, there is an activity peak at the 3:00am on Saturday which means in the late night of Friday, people take a transportation to go home]
  {\includegraphics[width=0.7\linewidth]{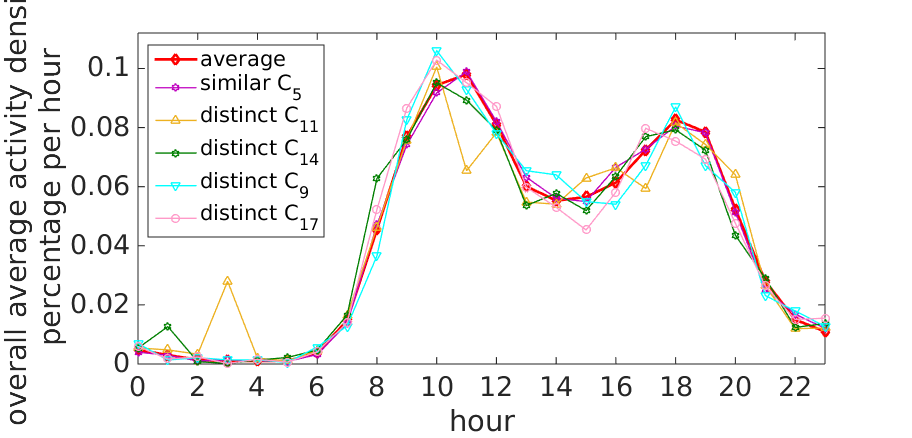}
  \label{gra:daily calling pattern saturday}}
  \quad
  \subfigure[Daily communication activity pattern per cluster over Sunday. In the pattern of cluster $C_{14}$, there is a peak every 1am that might be resulted from the traveling by transport and working activity]
  {\includegraphics[width=0.7\linewidth]{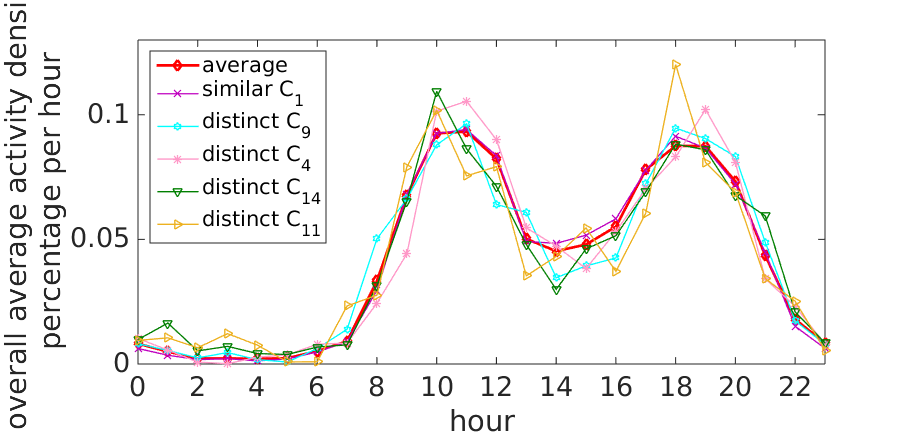}
  \label{gra:daily calling pattern sunday}}
  \caption{Typical communication activity temporal patterns over weekday vs weekend, the clusters $C_{11}$ and $C_{14}$ are the most distinct clusters compared to the average communication activity pattern over weekday, Saturday and Sunday}
\end{figure}

The daily communication activity timeline over weekday vs weekend (saturday and sunday) are shown in figures \ref{gra:daily calling pattern weekday}, \ref{gra:daily calling pattern saturday} and \ref{gra:daily calling pattern sunday}, respectively. They generally demonstrate quite similar pattern for different clusters. Based on the euclidean distance metric $C_{11}$ is described as the most distinct cluster pattern to the average communication activity pattern over weekday, Saturday and Sunday.

To assess the accuracy of the cluster quality when the ground truth of a dataset is not available, we have to use an intrinsic method. We evaluate these clusters by examining how well the clusters are separated and how compact the clusters are, based on the similarity metric (silhouette coefficient) between objects in the dataset: $a(o)=\frac{\sum_{o'\in C_i,o\neq o'}dist(o,o')}{\left | C_i  \right |-1}$, where $a(o)$ is the average distance between $o$ and all other objects in the cluster to which $o$ belongs. Similarly, $b(o)$ is the minimum average distance from $o$ to all clusters to which $o$ does not belong. Formally, suppose $o \in C_i (1 \leqslant i \leqslant k)$; then $b(o)=\min _{C_{j}:1\leq j\leq k,j\neq i} \left \{ \frac{\sum_{o'\in C_j}dist(o,o')}{\left | C_j  \right |} \right \}$. The silhouette coefficient is between $-$1 and 1, estimated by $s(o)=\frac{b(o)-a(o)}{\max\left \{ a(o),b(o) \right \}}$. The positive value reflects the more compactness of the cluster and well separated from other clusters. However, when the silhouette coefficient value is negative, the object in the considered same cluster is closer to the objects in another cluster. From the communication activity temporal patterns over the days of week, we estimated the silhouette coefficients of each area in all clusters based on the euclidean distance measure. This specific transport activity cluster $C_{11}$ is estimated with the clustering quality of 67\% where the silhouette coefficient of the objects in the cluster are positive. This means, the cluster is well separated from the other clusters and compact. This quality measure is increased to 77\% when we estimate the coefficient from the communication activity temporal patterns over whole time periods. We picked it up for further analysis as we expect traveling to be largely affected by special events and also as this cluster is a particular one (in terms of deviations of the timeline from average), at the same time having quality measure.

\begin{figure*}[h!]
  \centering
  \includegraphics[width=0.85\linewidth]{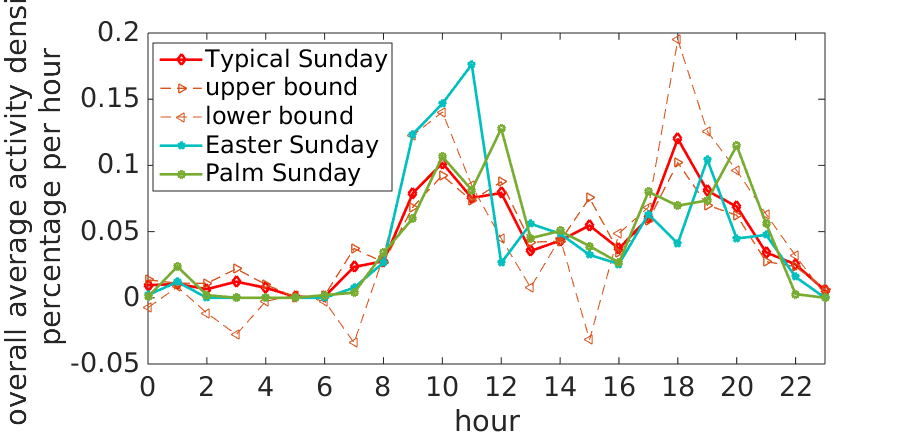}
  \caption{The daily temporal communication activity pattern of $C_{11}$ over typical Sunday compared with the communication activity pattern over Easter Sunday and Palm Sunday. There is an activity peak in the morning of Easter Sunday which is relatively increased than the typical variations of the communication activity, $\alpha=3$}
  \label{gra:calling pattern holiday}
\end{figure*}

We further investigate exceptional (divergent) type of temporal patterns over public events to determine how much the communication activity deviates from the variations of typical communication activities using the sigma approach. Figure \ref{gra:calling pattern holiday} shows the changes of communication activity temporal pattern over Easter Sunday and Palm Sunday. The Easter Sunday has a great impact of human behaviors as there is an activity peak in the morning of the Easter Sunday.

\section{Conclusion}\label{conclusion}

In this paper, we proposed an approach to identify relevant area clusters in terms of the categories of human activity (i.e., working, shopping or entertainment areas). The area clusters are used to contextualize mobile communication activities temporal patterns with the categories of human activity. An intrinsic method is used to assess the clustering quality if the cluster is well separated from other clusters and compact. And it turns out that communication activity, namely its density and temporal variation - is largely affected by the context of human activity in the area. The transport activity cluster $C_{11}$ is well classified with the clustering quality of 77\%. With the use of those area clusters, we explain the typical or exceptional (divergent) type of mobile communication activities. In future works, we evaluate the approach in different cities and measure the relation between the other types of human activity and mobile communication activities. The result of the research work is potentially useful for more coherent classifications of human behaviors and better understanding the relationship between human behaviors and environmental factors and their dynamics in real-life social phenomena.

\section{Acknowledgments}
The authors would like to thank the Semantic Innovation Knowledge Lab - Telecom Italia for providing the mobile phone data records. We also would like to thank MIT SENSEable City Lab Consortium for supporting the research.

\bibliographystyle{abbrv}
\bibliography{ref}

\end{document}